\documentclass[aps,prl,showpacs,showkeys,
twocolumn,superscriptaddress,altaffilletter,
amsmath,amssymb]{revtex4-1}
\usepackage{graphicx}
\usepackage{dcolumn}
\usepackage{bm}
\usepackage{upgreek}
\usepackage{color}
\usepackage{hyperref}
\usepackage{natbib}
\usepackage[mathlines]{lineno}
\newcommand{\ctsper}      {cts/(keV$\cdot$kg$\cdot$yr)}

\newcommand{\pIbi}        {{$10^{-2}$~cts/(keV$\cdot$kg$\cdot$yr)}}

\newcommand{\pIIbi  }     {{$10^{-3}$~cts/(keV$\cdot$kg$\cdot$yr)}}

\newcommand{\kgyr}        {{kg$\cdot$yr}}

\newcommand{\qbb}         {{$Q_{\beta\beta}$}}
\newcommand{\thalfzero}   {${T^{0\nu}_{1/2}}$}

\newcommand{\onbb}        {{$0\nu\beta\beta$}}

\newcommand{\gerda}       {\textsc{Gerda}}

\newcommand{\legend}       {\textsc{Legend}}
\newcommand{\mjd}         {\textsc{Majorana} Demonstrator}
\newcommand{\majorana}    {\textsc{Majorana}}

\newcommand{\gesix}       {{$^{76}$Ge}}
\newcommand{\gess}        {{$^{76}$Ge}}

\begin{document}
\title{
          Improved limit on  neutrinoless double beta
          decay of $^{76}$Ge from {\sc Gerda} Phase~II 
}
%
\collaboration{{\textsc{Gerda} collaboration}}
\email{correspondence: gerda-eb@mpi-hd.mpg.de}
\noaffiliation
%
  \affiliation{INFN Laboratori Nazionali del Gran Sasso and Gran Sasso Science Institute, Assergi, Italy}
  \affiliation{INFN Laboratori Nazionali del Gran Sasso and Universit{\`a} degli Studi dell'Aquila, L'Aquila, Italy}
  \affiliation{INFN Laboratori Nazionali del Sud, Catania, Italy}
  \affiliation{Institute of Physics, Jagiellonian University, Cracow, Poland}
  \affiliation{Institut f{\"u}r Kern- und Teilchenphysik, Technische Universit{\"a}t Dresden, Dresden, Germany}
  \affiliation{Joint Institute for Nuclear Research, Dubna, Russia}
  \affiliation{European Commission, JRC-Geel, Geel, Belgium}
  \affiliation{Max-Planck-Institut f{\"u}r Kernphysik, Heidelberg, Germany}
  \affiliation{Dipartimento di Fisica, Universit{\`a} Milano Bicocca, Milan, Italy}
  \affiliation{INFN Milano Bicocca, Milan, Italy}
  \affiliation{Dipartimento di Fisica, Universit{\`a} degli Studi di Milano e INFN Milano, Milan, Italy}
  \affiliation{Institute for Nuclear Research of the Russian Academy of Sciences, Moscow, Russia}
  \affiliation{Institute for Theoretical and Experimental Physics, NRC ``Kurchatov Institute'', Moscow, Russia}
  \affiliation{National Research Centre ``Kurchatov Institute'', Moscow, Russia}
  \affiliation{Max-Planck-Institut f{\"ur} Physik, Munich, Germany}
  \affiliation{Physik Department and Excellence Cluster Universe, Technische  Universit{\"a}t M{\"u}nchen, Germany}
  \affiliation{Dipartimento di Fisica e Astronomia dell{`}Universit{\`a} di Padova, Padua, Italy}
  \affiliation{INFN  Padova, Padua, Italy}
  \affiliation{Physikalisches Institut, Eberhard Karls Universit{\"a}t T{\"u}bingen, T{\"u}bingen, Germany}
  \affiliation{Physik Institut der Universit{\"a}t Z{\"u}rich, Z{u}rich, Switzerland}

\author{M.~Agostini}
  \altaffiliation[presently at: ]{Technische  Universit{\"a}t M{\"u}nchen, Germany}
  \affiliation{INFN Laboratori Nazionali del Gran Sasso and Gran Sasso Science Institute, Assergi, Italy}
\author{A.M.~Bakalyarov}
  \affiliation{National Research Centre ``Kurchatov Institute'', Moscow, Russia}
\author{M.~Balata}
  \affiliation{INFN Laboratori Nazionali del Gran Sasso and Gran Sasso Science Institute, Assergi, Italy}
\author{I.~Barabanov}
  \affiliation{Institute for Nuclear Research of the Russian Academy of Sciences, Moscow, Russia}
\author{L.~Baudis}
  \affiliation{Physik Institut der Universit{\"a}t Z{\"u}rich, Z{u}rich, Switzerland}
\author{C.~Bauer}
  \affiliation{Max-Planck-Institut f{\"u}r Kernphysik, Heidelberg, Germany}
\author{E.~Bellotti}
  \affiliation{Dipartimento di Fisica, Universit{\`a} Milano Bicocca, Milan, Italy}
  \affiliation{INFN Milano Bicocca, Milan, Italy}
\author{S.~Belogurov}
  \altaffiliation[also at: ]{NRNU MEPhI, Moscow, Russia}
  \affiliation{Institute for Theoretical and Experimental Physics, NRC ``Kurchatov Institute'', Moscow, Russia}
  \affiliation{Institute for Nuclear Research of the Russian Academy of Sciences, Moscow, Russia}
\author{A.~Bettini}
  \affiliation{Dipartimento di Fisica e Astronomia dell{`}Universit{\`a} di Padova, Padua, Italy}
  \affiliation{INFN  Padova, Padua, Italy}
\author{L.~Bezrukov}
  \affiliation{Institute for Nuclear Research of the Russian Academy of Sciences, Moscow, Russia}
% ...

\date{\today}

\begin{abstract}
 The \gerda\ experiment searches for the lepton number violating neutrinoless
 double beta decay of $^{76}$Ge
 ({$^{76}$Ge\,\,$\rightarrow$}\,$^{76}${Se}$\,+\,2e^-$) operating bare Ge
 diodes with an enriched $^{76}$Ge fraction in liquid argon. The exposure for
 BEGe-type detectors is increased threefold with respect to our previous data
 release. The BEGe detectors feature an excellent background suppression from
 the analysis of the time profile of the detector signals.  In the analysis
 window a background level of $1.0_{-0.4}^{+0.6}\cdot10^{-3}$\,\ctsper\ has
 been achieved; if normalized to the energy resolution this is the lowest ever
 achieved in any 0$\nu\beta\beta$ experiment.  No signal is observed and a new
 90\,\% C.L.~lower limit for the half-life of $8.0\cdot10^{25}$~yr is placed
 when combining with our previous data. The median expected sensitivity
 assuming no signal is $5.8\cdot10^{25}$\,yr.
\end{abstract}
\vfill
\maketitle

\section{Introduction}

 Despite many decades of research several properties of neutrinos are still
 unknown. Among them is the fundamental question whether neutrinos are their
 own anti-particles (i.e.~Majorana particles), as predicted by several
 extensions of the Standard Model of particle
 physics~\cite{mohapatra06,mohapatra07,rodejohann15}.  In this case
 neutrinoless double beta (0$\nu\beta\beta$) decay could be observed, a
 process in which lepton number is not conserved.

 Several experiments are taking data or are under preparation searching
 for this decay using a variety of suitable isotopes (see
 Refs.~\cite{Vergados:2016hso,limits:2017:matteo} for overviews). The sum of
 the kinetic energies of the two electrons emitted in the \onbb\ decay
 $(A,Z)\rightarrow(A,Z+2)+2\,e^-$ is equal to the mass difference \qbb\ of the
 two nuclei. A sharp peak in the energy spectrum is the prime signature for
 all \onbb\ experiments.

 Key parameters of these rare event searches are large mass $M$ and long
 measuring time $t$ on the one hand, and high energy resolution and low
 background on the other. Apart from the various isotopes the experiments
 differ in their setups and detection methods thereby exploiting the
 aforementioned parameters.  The GERmanium Detector Array (GERDA) experiment
 searches for \onbb\ decay of \gess\ using germanium detectors made from
 material enriched in \gess, i.e. source and detector are identical. This
 Letter shows that superior energy resolution and background suppression
 permit to achieve very sensitive results already at relative low exposure
 ${\cal E}=M{\cdot}t$.

\section{Experiment}

 The \gerda\ experiment is located at the Gran Sasso underground laboratory
 (LNGS) of INFN in Italy.  High-purity germanium detectors made from material
 with enriched \gess\ fraction of $\sim$87\,\% are operated in a 64~m$^3$
 liquid argon (LAr) bath. The argon cryostat is located inside a tank filled
 with 590~m$^3$ of high purity water. LAr and water shield against the
 external radioactivity. The water tank is instrumented with photomultipliers
 and operates as a Cherenkov detector to veto residual muon-induced events.
 Material for structural support of the detectors and for cabling is minimized
 in order to limit the background from close-by radioactive sources. More
 details of the experiment can be found in
 Refs.~\cite{gerda:2013:tec,gerda:2017:nature,gerda:2017:upgrade}.

 A first phase of data taking ended in 2013 with no indication of a
 signal~\cite{gerda:2013:prl}. The background index achieved at the
 \gess\ \qbb-value of 2039~keV was \pIbi.  For the second phase a new
 component has been installed to detect argon scintillation
 light~\cite{gerda:2017:upgrade}.  The enriched germanium mass was doubled in
 the form of small read-out electrode detectors (the Canberra BEGe detector
 model~\cite{gerda:2015:bege}) supplementing the previously used coaxial
 detectors.  Both enhancements allow for a more efficient rejection of
 background events, which can be characterized by their energy deposition in
 the LAr, in several detectors or in several locations (including the surface)
 of a single detector. In contrast, \onbb\ energy deposits are made by two
 electrons, which typically release all their energy in a small volume of a
 single detector.  Localized and delocalized energy deposits are distinguished
 by pulse shape discrimination (PSD) based on the time profile of the detector
 signal. For BEGe detectors a simple variable $A/E$ (the maximum $A$ of the
 detector current signal normalized by the energy $E$) shows very good PSD
 performance which is superior to the one based on neural networks for coaxial
 detectors~\cite{gerda:2013:psd}.

 Phase~II data taking started in December 2015 with a target background index
 of \pIIbi, a ten-fold reduction of background with respect to Phase~I. Thirty
 BEGe detectors (20.0~kg total mass) and seven coaxial detectors (15.6~kg) are
 deployed, whose energy resolution at \qbb\ is typically better than 3~keV and
 4~keV full width at half maximum (FWHM), respectively.
 
 As in Phase~I, a $\pm$25~keV window around \qbb\ was blinded: events
 with an energy in one detector within this window were hidden until the
 entire data selection was finalized. The first unblinding of Phase~II  took
 place in June 2016 and no \onbb\ signal was found. A lower limit of
 $T_{1/2}^{0\nu}>5.3\cdot10^{25}$~yr (90\,\% C.L.)  was extracted with a
 sensitivity, defined as the median expected lower limit assuming no
 signal, of $4.0\cdot10^{25}$~yr~\cite{gerda:2017:nature}.

\section{Results}

 Here, the result from a second unblinding of data from the BEGe detectors
 taken between June 2016 and April 2017 is reported.  The complete analysis of
 the new data set, including the detector energy reconstruction, LAr veto
 reconstruction, data selection, PSD and statistical treatment, is identical to
 the previous one published in Refs.~\cite{gerda:2017:nature,gerda:2013:prl}.

\begin{figure*}[tbp]
\begin{center}
\includegraphics[width=\textwidth]{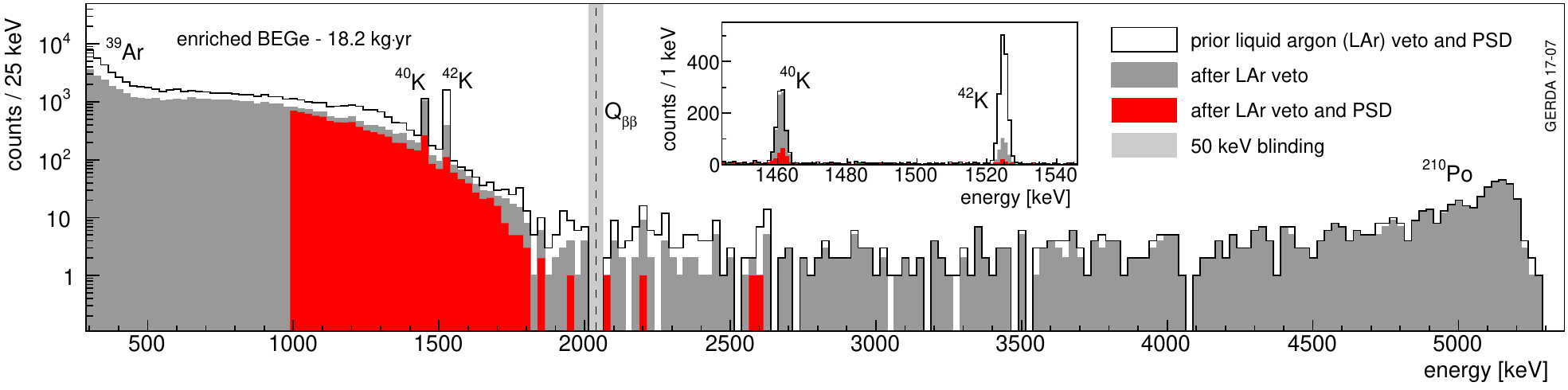}
\caption{\label{fig:bege}
    Energy spectra of Phase~II BEGe detectors prior to liquid argon (LAr) veto
    and PSD cuts (total histogram), after additional LAr veto (dark grey) and
    after after all cuts (red).  The inset shows the spectrum in the energy
    region of the potassium lines (1460~keV from $^{40}$K and 1525~keV from
    $^{42}$K). The grey vertical band indicates the blinded region of
    $\pm$25~keV around the \qbb-value.
}
\end{center}
\end{figure*}

With the new exposure of 12.4~\kgyr, the total Phase~II exposure doubles and
the one for the lower background BEGe detectors triples. Fig.~\ref{fig:bege}
shows the energy spectrum of the latter; the blinded region around \qbb\ is
indicated by the grey vertical band.  The spectrum below 500~keV is dominated
by $^{39}$Ar events, while the spectrum between 500 and 1800~keV is dominated
by events from $2\nu\beta\beta$ decays of $^{76}$Ge and Compton continua
mainly from the $^{40}$K and $^{42}$K lines. $\alpha$ decays dominate the
spectrum above 2620~keV. They are almost exclusively due to $^{210}$Po decays
at the p+ electrode or the isolating groove between p+ and n+ electrodes
(degraded $\alpha$ particles).  Since the $^{40}$K $\gamma$ line is from an
electron capture, no energy is deposited in the LAr and only PSD is effective
for rejecting events (see inset). The $\gamma$ line of $^{42}$K, the progeny
of the long-lived $^{42}$Ar, originates from a $\beta$ decay which deposits up
to 2~MeV in the argon.  The LAr veto rejects more than 80\,\% of these events
(see inset).

Near \qbb\ the spectrum is composed of degraded $\alpha$'s, $\beta$'s of
$^{42}$K decays at the detector surface, and Compton scattered $\gamma$'s from
$^{214}$Bi and $^{208}$Tl decays.  The background is evaluated in the range
between 1930 and 2190~keV without the two intervals (2104$\pm$5)~keV and
(2119$\pm$5)~keV from known $\gamma$ lines and without the signal interval
(\qbb$\pm$5)~keV. The analysis window for the \onbb\ search is identical but
includes the signal interval.  The low background index of BEGe detectors,
previously based on one single event, is now confirmed with a more than
threefold exposure to be $BI=1.0_{-0.4}^{+0.6}\cdot10^{-3}$~\ctsper. If
normalized according to the energy resolution and total signal
efficiency~$\epsilon$, i.e.~$BI$$\cdot$FWHM/$\epsilon$, this value corresponds
to 4.9$^{+2.9}_{-1.9}$~cts/(ton$\cdot$yr).  Hence, \gerda\ will remain
``background-free'', i.e. the average background in the energy interval
1$\cdot$FWHM at \qbb\ is expected to be less than~1 for the entire design
exposure of 100~\kgyr.  The efficiency $\epsilon$ (see Tab.~\ref{tab:tab1})
accounts globally for the abundance of \gess\ in the detectors, the active
volume fraction, the probability that the entire decay energy \qbb\ is
released in the active volume fraction of one Ge detector and the efficiency
of all selection and analysis cuts~\cite{gerda:2017:nature}.  The normalized
\gerda\ background $BI$$\cdot$FWHM/$\epsilon$ is at least a factor five lower
than that in any other competing non-\gess\ experiment.

The \mjd\ experiment also searches for \onbb\ decay of $^{76}$Ge employing
passive shielding made of ultra-pure copper. With the same
normalization~\footnote{
  \majorana\ reports a background rate of 4.0$^{+3.1}_{-2.5}$
  cts/(ton$\cdot$yr$\cdot$FWHM) \cite{mjd:2017}, which is normalized according
  to the active mass of the Ge detectors.  This has to be divided by the other
  efficiency factors reported in Ref.~\cite{mjd:2017}, namely (0.88$\pm$0.01)
  for \gesix\ enrichment and (0.80$\pm$0.03) for \onbb\ containment and
  selection
  efficiency.
}, their background is 5.7$^{+4.3}_{-3.6}$~cts/(ton$\cdot$yr). This result
is reported in the same issue of this journal~\cite{mjd:2017}.  Both
experiments have consequently extremely low background.

%
\begin{table}
  \caption{\label{tab:tab1}
    Summary of the Phase~I (PI) and Phase~II (PII) analysis datasets (exposure
    ${\cal E}$, energy resolution at \qbb\ (FWHM), total efficiency $\epsilon$
    and background index $BI$).
}
\begin{center}
\begin{tabular}{lcccc}
\hline\hline
{\rule{0mm}{4mm}}%
data set & ${\cal E}$ & FWHM  & $\epsilon$ & $BI$  \\
& [\kgyr] &  [keV] &  &
[10$^{-3}$\ctsper] \\[2mm]
\hline
{\rule{0mm}{4mm}}PI golden      & 17.9 & 4.3(1) & 0.57(3) & $11\pm2~$ \\
PI silver      & ~1.3 & 4.3(1) & 0.57(3) & $30\pm10$ \\
PI BEGe        & ~2.4 & 2.7(2) & 0.66(2) & $5^{+4}_{-3}$ \\
PI extra       & ~1.9 & 4.2(2) & 0.58(4) & $5^{+4}_{-3}$ \\
\quad total PI & 23.5 & & & \\
\hline
{\rule{0mm}{4mm}}PII coaxial    & ~5.0 & 4.0(2) & 0.53(5) & $3.5^{+2.1}_{-1.5}$ \\
PII BEGe       & 18.2 & 2.93(6)& 0.60(2) & $1.0^{+0.6}_{-0.4}$ \\
\quad total PII& 23.2 & & & \\
\hline
{\rule{0mm}{4mm}}\quad total  & 46.7 & & & \\
\hline\hline
\end{tabular}
\end{center}
\end{table}

The total exposure analyzed here is calculated from the total mass and amounts
to 23.5~\kgyr\ and 23.2~\kgyr\ for Phase~I and Phase~II, respectively. This
corresponds to (471.1$\pm$8.5)~mol$\cdot$yr of \gess\ in the active volume of
the detectors. Data from both phases are grouped in six data sets depending on
detector type and background level as summarized in Tab.~\ref{tab:tab1}.

\begin{table*}[ht]
 \caption{\label{tab:tab2}
       Comparison of lower half-life limits \thalfzero\ (90\,\% C.L.)  and
       corresponding upper Majorana neutrino mass $m_{\beta\beta}$ limits of
       different \onbb\ experiments. The experiments, the isotopes and the
       isotopic masses $M_i$ deployed are
       shown in cols. 1--3. The ranges of nuclear matrix elements (NME)
       \cite{mene09,horoi16,bar15,suh15,fae13,rod13,mus13,yao15} are given in
       col.~4. The lower half-life sensitivities and limits are shown in
       cols.~5 and~7, respectively.  The corresponding upper limits for 
       $m_{\beta\beta}$ derived with the NMEs are shown in cols. 6 and 8.
}
   \begin{tabular}{lrcrc|cc|cc}
\hline\hline
{\rule{0mm}{4mm}}%
&  &&&&\multicolumn{2}{c}{sensitivity}&\multicolumn{2}{|c}{limit} \\
   experiment && ~~~ isotope &  $M_i$ & NME &
      $T_{1/2}^{0\nu}$  & $m_{\beta\beta}$  &
                          $T_{1/2}^{0\nu}$  & $m_{\beta\beta}$\\
{\rule{0mm}{4mm}}  & & & [kg] & & [$10^{25}$~yr] & [eV] & [$10^{25}$~yr] & [eV]\\
 \hline
\gerda\ {\rule{0mm}{3mm}}  &
           & \gess\  &  31 & 2.8-6.1 ~& 5.8  & 0.14--0.30 & 8.0  & 0.12--0.26\\
\majorana\                 &\cite{mjd:2017}
           & \gess\  &  26 & 2.8-6.1 ~& 2.1  & 0.23--0.51 & 1.9  & 0.24--0.53\\
KamLAND-Zen                &\cite{kamland:2016}
           &$^{136}$Xe& 343 & 1.6-4.8 ~& 5.6  & 0.07--0.22 & 10.7 & 0.05--0.16\\
EXO                        &\cite{exo:2014,exo:2012}
           &$^{136}$Xe& 161 & 1.6-4.8 ~& 1.9  & 0.13--0.37 & 1.1  & 0.17--0.49\\
CUORE                      &\cite{cuore:2017,cuore:2016}
           &$^{130}$Te& 206 & 1.4-6.4 ~& 0.7  & 0.16--0.73 & 1.5  & 0.11--0.50\\
\hline\hline
\end{tabular}
\end{table*}

\begin{figure}[tbp]
\begin{center}
\includegraphics[height=0.8\columnwidth]{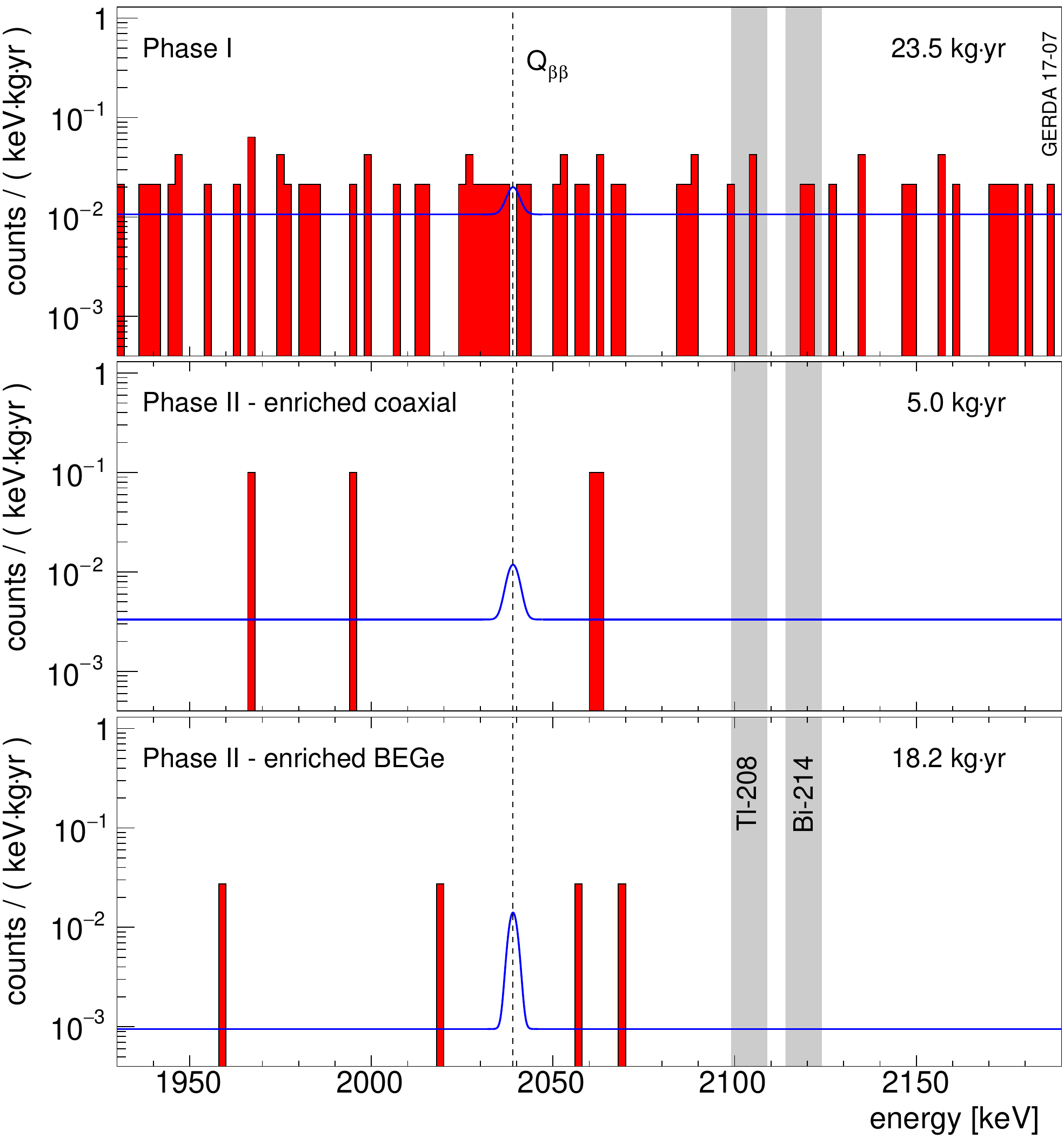}
\caption{\label{fig:fig2}
      Energy spectra in the analysis window for Phase~I and Phase~II coaxial
      detectors and Phase~II BEGe detectors, resp., after all cuts. The
      binning is 2~keV. The grey vertical bands indicate the intervals
      excluding known $\gamma$ lines. The blue lines show the hypothetical
      \onbb\ signal for \thalfzero$=8.0\cdot10^{25}$~yr, on top of their
      respective constant backgrounds.
}
\end{center}
\end{figure}
The spectrum in the analysis window is displayed in Fig.~\ref{fig:fig2}. Since
there is no event close to \qbb\ we place a 90\,\%~C.L.~lower limit of
$T_{1/2}^{0\nu}>8.0\cdot10^{25}$~yr on the decay half-life derived from a
frequentist (profile likelihood) analysis with a median sensitivity of
$5.8\cdot10^{25}$~yr.  The chance to have a stronger limit is 30\,\% as
evaluated by an ensemble of toy Monte Carlo realizations of the experiment
(for details of the statistical analysis see the `Methods' section in
Ref.~\cite{gerda:2017:nature}).  A Bayesian analysis with a flat prior in
$1/T_{1/2}^{0\nu}$ yields a lower limit of $5.1\cdot 10^{25}$~yr at 90\,\%
credibility and a sensitivity of $4.5\cdot 10^{25}$~yr.

\section{Discussion}

The lower half-life limit can be converted to an upper limit on the effective
Majorana neutrino mass $m_{\beta\beta}$ assuming the light neutrino exchange
as dominant mechanism.  Using the standard value of $g_A=1.27$, phase space
factors of Ref.~\cite{phasespace:2012:kotilla}, and the set of nuclear matrix
elements~\cite{mene09,horoi16,bar15,suh15,fae13,rod13,mus13,yao15} discussed
in a recent review~\cite{engel16}, the range for the upper limit on
$m_{\beta\beta}$ is 0.12--0.26 eV for \gesix.  The $m_{\beta\beta}$ limits for
several \onbb\ experiments obtained from profile likelihood analyses are
listed in Tab.~\ref{tab:tab2}. Despite the small deployed isotope mass $M_i$
the $m_{\beta\beta}$ sensitivity and actual limit of \gess\ are currently
merely a factor of $\approx$1.5 larger relative to the most sensitive one in
the field -- if the worst case NMEs are considered.

\gerda\ continues to collect data and is projected to reach a sensitivity on
the half-life well beyond $1\cdot10^{26}$~yr with the design exposure of
100~\kgyr.  The excellent energy resolution and extremely low background make
\gerda\ very well suited for a possible discovery, having a 50\,\% chance of a
$3\sigma$ evidence for a half-life up to $\sim$$8\cdot10^{25}$~yr at the
design exposure.

\gerda\ and \majorana\ Demonstrator both work in a ``background free''
regime. Therefore, the combined sensitivity on the \gesix\ \onbb\ decay will
increase almost linearly with the sum of the two exposures (see Fig.~2 of
Ref.~\cite{gerda:2017:upgrade}).  Having two experiments of similar background
obtained by different methods paves the way for the future
\legend\ experiment~\cite{legend:2017}.

\appendix
\section{Acknowledgments}       
 The \gerda\ experiment is supported financially by
   the German Federal Ministry for Education and Research (BMBF),
   the German Research Foundation (DFG) via the Excellence Cluster Universe
   and the SFB1258,
   the Italian Istituto Nazionale di Fisica Nucleare (INFN),
   the Max Planck Society (MPG),
   the Polish National Science Centre (NCN),
   the Foundation for Polish Science (TEAM/2016-2/2017),
   the Russian Foundation for Basic Research (RFBR), and
   the Swiss National Science Foundation (SNF).
 The institutions acknowledge also internal financial support.

This project has received funding/support from the European Union's Horizon
2020 research and innovation programme under the Marie Sklodowska-Curie grant
agreements No 690575 and No 674896, resp.

The \gerda\ collaboration thanks the directors and the staff of the LNGS
for their continuous strong support of the \gerda\ experiment.
\providecommand{\noopsort}[1]{}\providecommand{\singleletter}[1]{#1}%

\end{document}